\documentclass[aps,pra,twocolumn,groupedaddress,showpacs,floatfix]{revtex4}
\usepackage{amssymb}
\usepackage{graphicx}
\usepackage{subfigure}	 	
\usepackage[english]{babel}
\usepackage{float}

\begin{document}

\title{Multi-stable regime and intermediate solutions in a nonlinear saturable coupler}

\author{Diego Guzm\'an-Silva$^{1}$, Cibo Lou$^{2}$, Uta\ Naether$^{1}$, Christian E. R\"uter$^{2}$, Detlef Kip$^{2}$ and Rodrigo A. Vicencio$^{1}$}
\affiliation{$^1$Departamento de F\'{\i}sica, MSI-Nucleus on Advanced Optics, and Center for Optics and Photonics (CEFOP), Facultad de Ciencias, Universidad de Chile, Santiago, Chile}
\affiliation{$^2$Faculty of Electrical Engineering, Helmut Schmidt University, 22043 Hamburg, Germany}
\date{\today}

\begin{abstract} 

We show, theoretically and experimentally, the existence of a multi-stable regime in a nonlinear saturable coupler. In spite of its simplicity, we found that this model shows generic and fundamental properties of extended saturable lattices. The study of this basic unit becomes crucial to understand localization mechanisms and dynamical properties of extended discrete nonlinear saturable systems. We theoretically predict the regions of existence of intermediate solutions, and experimentally confirm it by observing a multi-stable propagation regime in a LiNbO$_3$ saturable coupler. This constitutes the first experimental evidence of the existence of these unstable symmetry-broken stationary solutions. 

\end{abstract}


%\twocolumn[ %% activate for two-column option
\maketitle

\section{Introduction}

Nonlinear discrete systems appear in several branches of science and have found a fruitful field of development and realistic implementation during the last years~\cite{pt,rep1,rep2,rep3}. Many results obtained in very different physical settings can be extrapolated to other areas of research generating a broader and deeper scientific impact. Different techniques and methods to study such systems experimentally, for example in the context of photonic lattices, have been developed with many possibilities to change and control the key parameters. Here, merely Kerr-like (cubic) nonlinear systems have been a main subject of theoretical and experimental research. As a result, the corroboration of several former theoretical predictions and different new findings have been performed. However, a different type of nonlinearity has opened new challenges with new interesting dynamical properties, the so-called ``saturable nonlinearity''. On one hand, from a dynamical point of view, this type of nonlinearity allows for a more complex and richer phenomenology than typical cubic systems~\cite{prlkip, prlmel,1d,2da,2db}. For example, in a saturable nonlinear regime an exchange of stability properties between fundamental solutions is allowed, promoting improved mobility for high power solitons. This is certainly opposite to the phenomenology observed for cubic systems~\cite{moraprl,ol1}, where an increment in power induces localization only. On the other hand, only very little experimental realizations showing the specific behavior of saturable systems have been performed. So far, mainly gradual changes from the behavior of Kerr-like systems were observed; for example, suppression of modulation instability~\cite{x1}, stabilization of discrete vector solitons~\cite{x2}, and higher-order gap solitons~\cite{x3}.  

A nonlinear coupler (dimer) is the simplest discrete system where there are just two identical nonlinear waveguides (sites) - placed in close proximity - which evanescently interact. Thirty years ago, Jensen~\cite{jensen} showed the main features for the corresponding cubic system; i.e., a periodic exchange of light for small power and high transmission (localization) for larger powers. In Ref.~\cite{dimcub} authors explored the dynamics and stationary behavior of a cubic coupler which presents only one bifurcation point for stationary solutions and no exchange of stability properties [similar to larger one-dimensional (1D) cubic systems]. On the other hand, Ref.~\cite{dimsat} shows the appearance of an extra bifurcation point, change of stability properties and richer dynamics when considering a saturable nonlinearity. However, only very recently, the concept of ``intermediate solutions'' (IS) was introduced to explain the properties of extended saturable 1D and 2D lattices~\cite{1d,2da,2db} (this concept was introduced before for other models~\cite{ISs,mika}). These kind of ``unstable'' symmetry-broken solutions appear when two fundamental modes are simultaneously stable (in other settings, the IS can also be stable~\cite{mika,Abd08,dipolar}). Therefore, it becomes natural to formulate the question about the minimum number of sites - in a saturable array - for which this phenomenology emerges. In the present work, we will show that the fundamental saturable properties are already present in a system of just two waveguides. Moreover, for the first time to our knowledge, we show an evidence for the existence of IS by observing a multi-stable propagation in an experiment performed in iron doped LiNbO$_3$ samples.

\section{Model}

The propagation of light in a system composed of two identical weakly coupled waveguides, with a defocusing saturable nonlinearity, can be described as follows
\begin{eqnarray}
-i \frac{\partial u_{1}}{\partial z}=u_{2}+ \frac{\gamma u_{1}}{1+|u_{1}|^2}\ ,~~ 
-i \frac{\partial u_{2}}{\partial z}=u_{1}+ \frac{\gamma u_{2}}{1+|u_{2}|^2}\label{eq}
\end{eqnarray}
where $u_n$ represents the light amplitude at site $n$ , $\gamma\equiv \bar{\gamma}/V>0$ corresponds to the strength of the defocusing nonlinearity ($\bar{\gamma}$) with respect to the coupling coefficient ($V$) between the two sites, and $z$ describes the normalized propagation distance along the waveguides. Model (\ref{eq}) possesses two conserved quantities, the Power
\begin{equation}
P\equiv|u_1|^2+|u_2|^2
\label{P}
\end{equation}
and the Hamiltonian
\begin{equation}
H\equiv(u_{2}u^*_{1}+u_{2}^* u_{1})+\gamma \ln\left[(1+|u_{1}|^2)(1+|u_{2}|^2)\right]\ .
\label{H}
\end{equation}
Stationary solutions of model (\ref{eq}) have the form $u_n(z)=u_n \exp(i\lambda z)$, where $\lambda$ represents  the spatial frequency. First of all, we look for linear solutions of model (\ref{eq}) that, due to the saturation, exist in two different regions~\cite{1d,2da,2db}. In a low power regime, we find that there are two solutions with frequencies ``$\gamma+1$'' and ``$\gamma-1$''. For higher powers, the nonlinear response vanishes and frequencies become ``$+1$'' and ``$-1$''. The spatial profiles for these two modes are equal in both regimes: the symmetric ($u_1=u_2$) and the antisymmetric ($u_1=-u_2$) modes, respectively. 

\section{Nonlinear modes}

Now, we look for general nonlinear solutions of the form: $u_1=A$ and $u_2=\alpha A$, where $A$ is a positive amplitude and $\alpha$ describes the ratio between these two site amplitudes. We found that the symmetric ($\alpha=1$) and the antisymmetric ($\alpha=-1$) modes are also solutions in the nonlinear regime. They bifurcate from the low power linear solutions and diverge when approaching the high power linear modes:
\[
P_{sym}=2\left[\frac{(\gamma+1)-\lambda}{\lambda-1}\right]\ \  \text{and}\ \ P_{ant}=2\left[\frac{(\gamma-1)-\lambda}{\lambda+1}\right].
\]

\begin{figure}[htbp]
\centering
\includegraphics[width=0.47\textwidth]{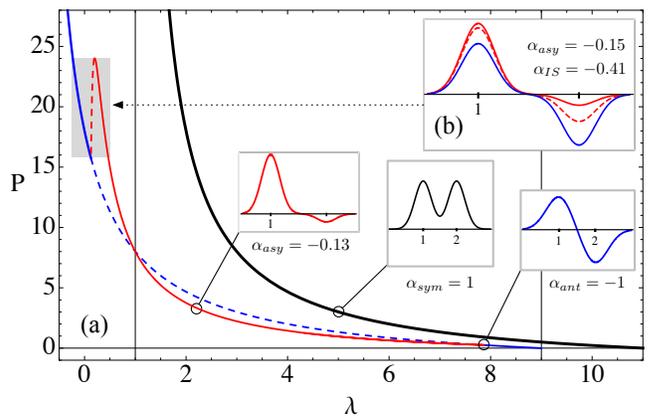}
\caption{(Color Online) (a) $P$ vs $\lambda$ diagram for $\gamma=10$. The symmetric, stable and unstable antisymmetric, asymmetric, and intermediate solutions are plotted in black, full and dashed blue lines, red, and red-dashed line, respectively. Vertical lines indicate linear frequencies. (b) Antisymmetric, asymmetric, and intermediate mode profiles, for $P=20$.}
\label{PL}
\end{figure}

In Fig.~\ref{PL}(a) these two families are plotted with thick-black and blue lines, respectively, including their corresponding profiles (to visualize them, we used a Gaussian profile with amplitude $u_n$ at site $n$). In addition, we found two non-symmetric solutions
\begin{equation}
\alpha(\gamma,A)=\frac{-\gamma A^2\pm\sqrt{\gamma^2 A^4-4A^2(1+A^2)^2}}{2A^2(1+A^2)}\ .
\label{alf}
\end{equation}
%
%
%%%%%%%%%%%%%%%%%%%%%%%%%%%%%%%%%%%%%%%%%%%%
%%%%%%%%%%%%%%%%%%%%%%%%%%%%%%%%%%%%%%%%%%%%
%
%
Without loss of generality, we restrict to the case $|\alpha|<1$. The sign ``$+$'' in (\ref{alf}) corresponds to the asymmetric solution ($|\alpha_{asy}|\neq 1$) that exists only for $\gamma>4$, bifurcating from the antisymmetric mode [see Fig.~\ref{PL}(a) at $\lambda\sim 8$]. Once this asymmetric solution appears, the antisymmetric one becomes unstable [a standard linear stability analysis~\cite{sta} was performed and full (dashed) lines indicate stable (unstable) solutions in Fig.~\ref{PL}]. Fig.~\ref{PL}(a) shows a monotonic increment of the asymmetric power up to some maximum value. All this branch (full red line) is stable. However, after achieving this maximum, the branch changes its curvature (power decreases) and the solution gets unstable until it fuses with the antisymmetric branch [this unstable branch corresponds to the sign ``$-$'' in Eq.~(\ref{alf})]. For saturable systems, this always unstable non-symmetric solution is called intermediate solution~\cite{2da,2db,1d}. 
The case where two fundamental solutions are simultaneously stable, sharing the same Hamiltonian value, was initially suggested~\cite{prlkip} as a vanishing Peierls-Nabarro barrier~\cite{pey,pnp}. However, recent works~\cite{mika,2da,2db,1d} have shown that there is a nonzero effective energy barrier, which also considers the IS. Surprisingly, the dimer model also shows this phenomenology which is fundamental to understand deeply the properties of nonlinear saturable arrays. Examples for some profiles are sketched in Fig.~\ref{PL}(b). The intermediate and asymmetric modes correspond both to non-symmetric solutions of model (\ref{eq}); they have quite similar profiles and would be only identified by directly observing their unstable/stable dynamical propagation.

\subsection{Effective potential}

Now, we go deeper into the dynamical properties of this model by computing an effective potential~\cite{surface1da,surface1db,2da,2db,1d}. Model (\ref{eq}) is integrable, therefore the effective potential can be obtained analytically. First of all, by fixing the Power $P$, we define the center of mass as $x\equiv u_2^2/P$ ($x=0$ or $1\Rightarrow P=u_1^2$ or $u_2^2$, and $x=0.5\Rightarrow u_1^2=u_2^2=P/2$). 
Then, as we are considering a defocusing nonlinearity, we study staggered solutions ($\alpha<0$) and express the amplitudes as
\[
u_1=\pm\sqrt{P(1-x)}\ \ \ \text{and}\ \ \ u_2=\mp\sqrt{xP}\ .
\]
With these expressions inserted into the Hamiltonian (\ref{H}), we get
\[
H(x,P,\gamma)=-2P\sqrt{x-x^2}+\gamma\ln \left [1+P+P^2(x-x^2)\right ].
\]
The critical points $\partial H/\partial x=0$ represent different stationary solutions. A first one corresponds to $x_{ant}\equiv0.5$ ($\alpha=-1$), a solution existing for all level of powers $P$ [see blue line in Fig.~\ref{PL}(a)]. There are four additional solutions
\begin{equation}
x=\frac{1}{2}\pm\frac{1}{2P}\sqrt{P^2-2\gamma^2+4(P+1)\pm 2\gamma\sqrt{\gamma^2-4(P+1)}},
\label{xes}
\end{equation}
where asymmetric modes correspond to the sign ``$+$'' in front to the inner square root. They are symmetrically located - to the right and to left - from the antisymmetric solution ($x_{ant}=1/2$). Asymmetric solutions exist only for $\gamma>4$, bifurcating from the antisymmetric mode at power $P_{min}\equiv\gamma-2-\sqrt{(\gamma-2)^2-4}$. Unstable IS correspond to the sign ``$-$'' in front to the inner square root in (\ref{xes}). They exist above the power threshold $P_{th}\equiv\gamma-2+\sqrt{(\gamma-2)^2-4}$. The asymmetric and the intermediate solutions exist up to an upper power value given by $P_{up}\equiv(\gamma/2)^2-1$. (For $\gamma=10$, $P_{min}=0.254$, $P_{th}=15.746$, and $P_{up}=24$ [see Fig.~\ref{PL}(a)]). IS also exist for $\gamma>4$; i.e, the intrinsic saturable phenomenology would be only observed above some critical nonlinearity.
%
%%%%%%%%%%%%%%%%%%%%%%%%%%%%%%%%%%%%%%%%%%%%
\begin{figure}[t]
\centering
\includegraphics[width=0.47\textwidth]{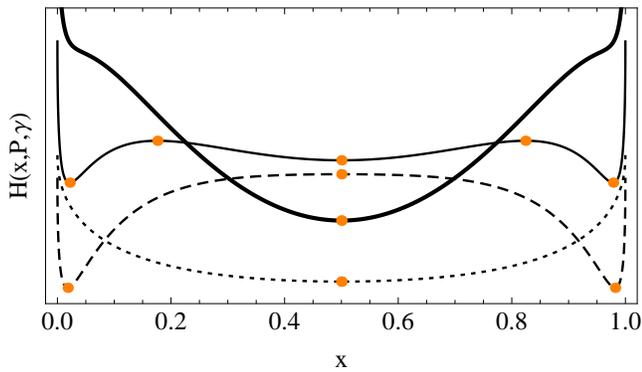}
\caption{(Color Online) Effective potential for $\gamma=10$ and $P=0.15$ (dotted), $P=15$ (dashed), $P=19$ (full) and $P=26$ (full thick). Filled circles correspond to stationary solutions.}
\label{HX}
\end{figure}
%%%%%%%%%%%%%%%%%%%%%%%%%%%%%%%%%%%%%%%%%%%%
%

Fig.~\ref{HX} shows the effective potential $H(x,P,\gamma)$ versus the center of mass, for $\gamma =10$ and for four different level of power (curves have been normalized for comparison). Below the $P_{min}$ (dotted line), the potential looks as a typical potential well with a minimum located at $x_{ant}$. In the range $\{P_{min},P_{th}\}$ the effective potential is cubic-like (dashed line); i.e, the antisymmetric solution is unstable (maximum) while the asymmetric solution is stable (minimum). The saturable nature of this model manifests for powers above $P_{th}$ (full line). The antisymmetric and the asymmetric solutions become, both, a local minimum and, therefore, simultaneously stable. The unstable symmetry-broken IS appears as a maximum in this potential, located in between the stable solutions. Therefore, at $P_{th}$ the IS has a center of mass $x_{IS}=x_{ant}$ that evolves in the direction of an asymmetric configuration when approaching $P_{up}$. Finally, above $P_{up}$, both non-symmetric solutions disappear, and the antisymmetric mode is the only critical point in the effective potential (full thick line). This is a direct consequence of the saturable nature of the system, where the nonlinear term vanishes for powers above a critical value ($P>P_{up}$).

\subsection{Numerical propagation}

\begin{figure}[t]
\centering
\includegraphics[width=0.47\textwidth]{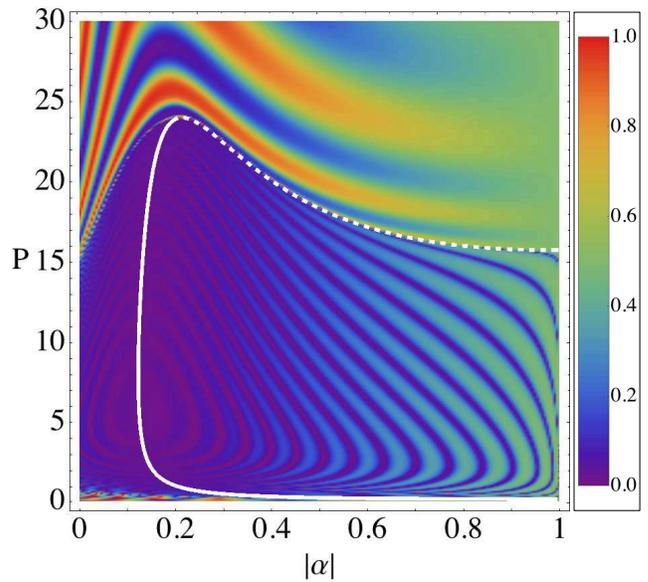}
\caption{(Color Online) Density Plot of ``$x_{out}$'' versus $|\alpha|$ and $P$ for $\gamma=10$. Full and dashed white lines correspond to the asymmetric and intermediate solutions, respectively.}
\label{alfa}
\end{figure}

Now, in order to test these stationary properties and their dynamical consequences, we numerically study model (\ref{eq}) by considering the general initial condition $u_1(0)=A$ and $u_2(0)=\alpha A$, with $-1\leqslant\alpha\leqslant0$.
This input condition allow us to excite all different staggered solutions. Fig.~\ref{alfa} shows the output center of mass, defined as $x_{out}\equiv|u_2(z_{max})|^2/P$, measured after a given propagation distance $z=z_{max}$ (similar density maps are obtained for different propagation distances).
%
%%%%%%%%%%%%%%%%%%%%%%%%%%%%%%%%%%%%%%%%%%%%
%%%%%%%%%%%%%%%%%%%%%%%%%%%%%%%%%%%%%%%%%%%%
%
Purple-blue colors ($x_{out}\lesssim 0.2$) represent solutions localized close to site $n=1$; green colors ($0.4\lesssim x_{out}\lesssim 0.6$) represent antisymmetric profiles; orange-red colors ($x_{out}\gtrsim 0.8$) represent solutions localized close to site $n=2$. For $P\lesssim 1$, asymmetric profiles ($|\alpha|\sim 0$) do not correspond to any stationary solution and the light just oscillates between the two sites (see the appearance of multiple colors as an indication of strong oscillation, i.e. switching~\cite{jensen}). On the other hand, for $|\alpha|\sim 1$ we see a greener color that indicates a small oscillation in the vicinity of the antisymmetric solution ($x_{out}=0.5$), the only stationary solution at this power regime. At the bifurcation point power (when $H$ changes its shape from dotted to dashed in Fig.~\ref{HX}), both solutions are quite similar and the antisymmetric mode is slightly unstable. Then, in the range $P\in\{1,15\}$ for $|\alpha|\lesssim 0.4$, the light is well trapped at the vicinity of site $n=1$. This is an indication of the excitation of an asymmetric stationary state (see full white line in Fig.~\ref{alfa}). For $|\alpha|\rightarrow 1$, there is an oscillation of the light in the interval $x_{out}\in\{0,0.5\}$, as expected for an unstable antisymmetric configuration (see dashed line in Fig.~\ref{HX}). In the region of power $\approx\{16,24\}$, a very interesting behavior is observed: we found a simultaneously stable dynamical propagation of asymmetric ($|\alpha|\approx 0.15$) and antisymmetric ($|\alpha|\approx 1$) input profiles. The constant value of $x$ (constant color) when increasing $P$, indicates a stable dynamical evolution. This constitutes a dynamical and indirect evidence of the existence of the intermediate solutions as a fundamental entity for saturable systems: \textit{if both staggered solutions are stable simultaneously, an extra unstable intermediate solution must exist}. The dashed white line in Fig.~\ref{alfa} corresponds to the Intermediate solution and shows a clear connection in parameter space between the asymmetric and the antisymmetric staggered solutions. For $P\gtrsim 24$, we observe an oscillation of energy between sites $1$ and $2$ for the input condition $0\lesssim|\alpha|\lesssim 0.8$, while an antisymmetric configuration is stable for $|\alpha|\gtrsim 0.8$. All the dynamical results are in perfect agreement with the stationary picture sketched in Figs.~\ref{PL} and \ref{HX}, including the region of multi-stability and the existence of the IS.

\section{Experimental results}

%%%%%%%%%%%%%%%%%%
\begin{figure}[h]
\includegraphics[width=0.47\textwidth]{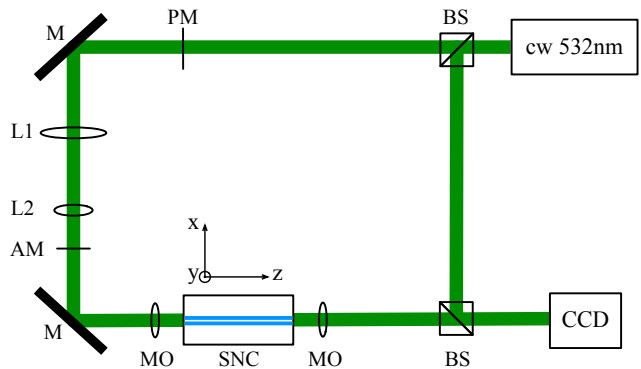}
\caption{(Color Online) Experimental setup for observing the multistable regime in a saturable nonlinear coupler.}\label{setup}
\end{figure}
\begin{figure}[h]
\includegraphics[width=8.6cm]{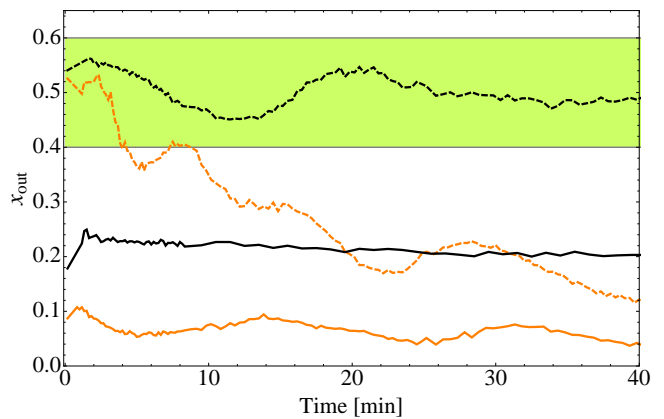}
\caption{(Color Online) Center of mass evolution at the sample output for asymmetric (full lines) and antisymmetric (dashed lines) input conditions, for input powers $100$nW (black) and $500$nW (orange). The filled rectangular area denotes the region where we experimentally observe (define) an antisymmetric configuration.}
\label{data0}
\end{figure}

%%%%%%%%%%%%%%
%
To verify our theoretical and numerical predictions we use the experimental setup sketched in Fig.~\ref{setup}. A cw laser with a wavelength of $532$ nm propagates through a phase mask (PM) covering half of the beam along the transverse direction $x$ (coinciding with the crystallographic c-axis of the sample). So the phase relation of the left and right half of the beam can be switched to be either in-phase or out-of-phase. With a 4f imaging system, composed by lenses L1 and L2, the beam is imaged onto a double-hole amplitude mask (AM). By using a microscope objective, the beam is injected into a saturable nonlinear coupler (SNC) fabricated by titanium in-diffusion on an $x$-cut lithium niobate substrate doped with iron. The end facet of the sample is monitored by a high-resolution CCD camera. The length of our sample along the propagation $z$-direction is 18 mm ($\sim 5$ diffraction lengths) where the waveguide channels are 4.0 $\mu$m wide with a separation of 2.2 $\mu$m. Our photovoltaic samples have a nonlinearity which grows exponentially in time, $\gamma(t) =\gamma(1-\exp[-t/\tau])$, where $\tau$ is the dielectric response time~\cite{gama}. In order to reach a steady-state, saturation of the photovoltaic nonlinearity is required, which typically occurs, in our samples, for $t\sim 25$ minutes. Fig.~\ref{data0} shows some examples of the evolution of the center of mass in time. We see how some input conditions propagates in a very stable way for the whole measurement period (black-full, orange-full and black-dashed lines) while the orange-dashed curve tend to stabilize for $t\gtrsim 25$.
%
%%%%%%
%%%%%%%%
%

\begin{figure}[t]
\includegraphics[width=8.6cm]{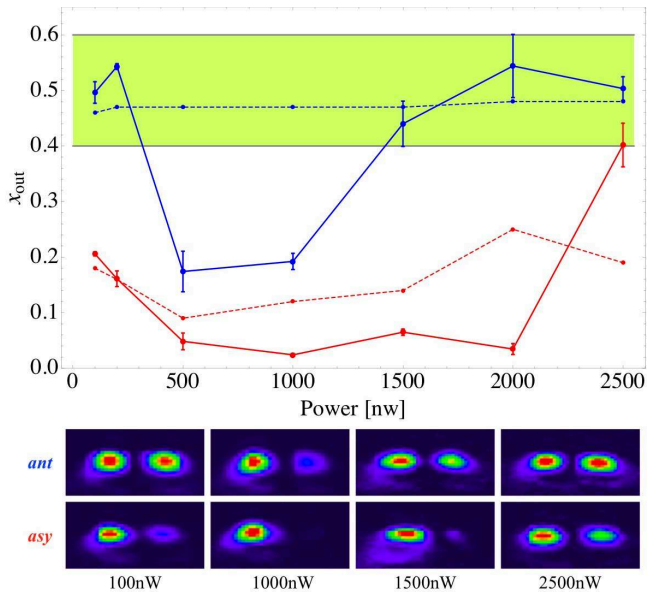}
\caption{(Color Online) Averaged center of mass at the sample output vs input power for a LiNbO$_3$ saturable coupler. Red and blue symbols, connected by full lines, correspond to the experimentally measured value of $x_{out}$ for an asymmetric and antisymmetric input condition (symbols connected by dashed lines), respectively. The lower row shows some (non-averaged) experimental output profiles for the indicated powers with antisymmetric (ant) and asymmetric (asy) input profiles. (The filled area denotes the same that in Fig.~\ref{data0}).}
\label{data}
\end{figure}

We repeat the experiment for several input powers and for the asymmetric and the antisymmetric input profiles, and average the center of mass values for the last $10$ minutes of each experiment. Compiled results are shown in Fig.~\ref{data} with a fairly good agreement between the theoretically predicted phenomenology and what is observed in direct experiments. The initial conditions are indicated by dots connected with dashed lines. The deviation of the output values (dots connected by full lines) from the initial conditions can be understood as the degree of stability. In addition, the error bars (obtained in the averaging process) also indicates how stable is the profile: a smaller bar indicates a dynamically stable profile while a larger bar means a stronger oscillation around some minimum (stationary solution). For low level of power, we observe that both solutions are essentially stable. This coincides with our analysis in the region of the first bifurcation point, where the appearance of the asymmetric solution weakly destabilizes the antisymmetric mode. Then, for powers in the region $\sim\{300,1300\}$nW, a typical cubic-like picture is observed: the asymmetric solution is stable while the antisymmetric one is not. However, above $\sim 1300$nW a saturable phenomenology emerges. The saturable nature of the nonlinearity allows oscillation of the stability properties, with a richer dynamics when comparing with usual cubic systems. For this sample, the region $\sim\{1300,2300\}$nW corresponds to a regime where the two fundamental nonlinear solutions of this problem become simultaneously stable (as an example see the profiles for $1500$nW). Therefore, we experimentally observe a multi-stable regime as an indirect evidence of the existence of intermediate solutions in a saturable dimer. This observation gives a strong support to the theory (and model) developed for this type of lattices. Above some given power ($\sim 2300$nW), the only stable solution is the antisymmetric one, as shown in Fig.~\ref{data}. The large bar for the last asymmetric (red) point indicates that the system has saturated and thadit the effective potential has a shape like the thick line in Fig.~\ref{HX}. It is important to mention that our experimental output profiles in Fig.~\ref{data}-low-row have a staggered phase, what is evident by observing the zero amplitude in between the two waveguides. This gives us an extra support for relating the observation of the multi-stable regime with the existence of an unstable IS, because the observed profiles are well connected in phase space.

%
%%%%%%
%%%%%%%%
%

\section{Conclusions}

In conclusion, we have observed, for the first time, a multi-stable regime of fundamental modes in a nonlinear saturable coupler. A complete map of nonlinear solutions has been constructed, including their stability properties, and effective potential. Numerically, we have determined the regions where the fundamental profiles are expected to be stable, showing an excellent agreement with the developed theory. We fabricated a nonlinear saturable coupler in LiNbO$_3$ and observed stable propagation of fundamental modes for intermediate level of power. This constitutes the first experimental evidence, in any physical system, of the existence of intermediate solutions in discrete nonlinear lattices. Moreover, for larger powers we observed the absence of the asymmetric solution as a clear indication of the saturation of the nonlinearity. All these results support strongly recent theoretical and numerical developments based on DNLS-like models in several contexts of physics and open, as a direct consequence, new opportunities of research on nonlinear discrete systems. In addition, the understanding of small systems becomes very crucial when thinking on the implementation of photonic lattices for realistic applications. 

\begin{acknowledgements}
The authors thank M. Johansson for useful discussions. This work was supported in part by FONDECYT grant 1110142, CONICYT fellowships, Programa ICM P10-030-F, Programa de Financiamiento Basal de CONICYT (FB0824/2008), and the Deutsche Forschungsgemeinschaft (Ki 482/14-1).
\end{acknowledgements}

\end{document}